\documentclass[a4paper, 12pt]{article}  
\usepackage{pdfpages}
\usepackage{lmodern}
\usepackage{graphicx} 
\usepackage{wrapfig} 
\usepackage{textcomp} 
\usepackage[utf8]{inputenc}
\DeclareSymbolFont{letters}{OML}{ztmcm}{m}{it}
\usepackage{color}
\usepackage{bm}
\usepackage{multirow}

\usepackage{array}
\usepackage{listings}
\usepackage{adjustbox}
\usepackage[boxed]{algorithm2e}
\usepackage{subcaption}
\usepackage{multirow} 
\usepackage[OMLmathrm,OMLmathbf]{isomath}
\usepackage{booktabs} 
\usepackage[hang,flushmargin]{footmisc}
\usepackage{pdflscape}
\usepackage{natbib}

\usepackage[T1]{fontenc} 
\usepackage{geometry}
\usepackage{amsmath}
\usepackage{etoolbox} 
\usepackage{amsthm}
\usepackage{amsfonts}
\usepackage{amssymb}
\usepackage[section]{placeins}
\usepackage{diagbox} 
\usepackage{hyperref}
\usepackage{graphicx} 
\usepackage{fixmath} 
\usepackage[scaled=.90]{helvet}
\usepackage{wrapfig}
\usepackage{tikz}
\usepackage{footnote}
\makesavenoteenv{tabular}
\makesavenoteenv{table}
\usepackage{setspace}
\usepackage[boxed]{algorithm2e}

\makeatletter
\renewcommand\@biblabel[1]{\textbf{#1.}} 
\usepackage{enumitem}
\newcommand{\blind}{1}
\usepackage{pifont}
\setlist{nolistsep,leftmargin=*}
\renewcommand{\maketitle}{ 
	\begin{center}
		{\LARGE\@title} 
		
		\vspace{0pt} 
		
		{\large\@author} 
		\\\@date 
		
		\vspace{40pt} 
	\end{center}
}
\begin{document}
	
	\def\spacingset#1{\renewcommand{\baselinestretch}%
		{#1}\small\normalsize} \spacingset{1}

	
	\if1\blind
	{
	
		\title{\textbf{All that Glitters is not Gold: \\  Relational Events Models with Spurious Events\\}}
		\author{Cornelius Fritz$^\ast$\footnote{Corresponding Autor: \href{mailto:cornelius.fritz@stat.uni-muenchen.de}{cornelius.fritz@stat.uni-muenchen.de}}, Marius Mehrl$^\dagger$, Paul W. Thurner$^\dagger$, Göran Kauermann$^\ast$\hspace{.2cm}\\
	Department of Statistics, LMU Munich$^\ast$\hspace{.2cm}\\ Geschwister Scholl Institute of Political Science, LMU Munich$^\dagger$ 
	 }
		\maketitle
	} \fi

	\if0\blind
	{
		\bigskip
		\bigskip
		\bigskip
		\begin{center}
			{\LARGE\bf Relational Events with Measurement Errors}
		\end{center}
		\medskip
	} \fi
	
	\bigskip

\begin{abstract}
   As relational event models are an increasingly popular model for studying relational structures, the reliability of large-scale event data collection becomes more and more important. Automated or human-coded events often suffer from non-negligible false-discovery rates in event identification. And most sensor data is primarily based on actors' spatial proximity for predefined time windows; hence, the observed events could relate either to a social relationship or random co-location. Both examples imply spurious events that may bias estimates and inference. We propose the Relational Event Model for Spurious Events (REMSE), an extension to existing approaches for interaction data. The model provides a flexible solution for modeling data while controlling for spurious events. Estimation of our model is carried out in an empirical Bayesian approach via data augmentation. Based on a simulation study, we investigate the properties of the estimation procedure. To demonstrate its usefulness in two distinct applications, we employ this model to combat events from the Syrian civil war and student co-location data. Results from the simulation and the applications identify the REMSE as a suitable approach to modeling relational event data in the presence of spurious events.     
\end{abstract}


\section{Introduction}


In recent years, event data have become ubiquitous in the social sciences. For instance, interpersonal structures are examined using face-to-face interactions \citep{Elmer2020}. At the same time, political event data are employed to study and predict the occurrence and intensity of armed conflict \citep{Fjelde_Hultman_2014,Blair2020,Dorff_Gallop_Minhas_2020}. \citet{butts2008} introduced the Relational Event Model (REM) to study such relational event data. In comparison to standard network data of durable relations observed at specific time points, relational events describe instantaneous actions or, put differently, interactions at a fine-grained temporal resolution \citep{Borgatti2009}. 

However, in some contexts there arise problems regarding the reliability of event data. While data gathered from e.g. direct observations \citep{tranmer2015} or parliamentary records \citep{Malang2019} should prove unproblematic in this regard, other data collection methods may be prone to \textsl{spurious events}, i.e. events that are recorded but did not actually occur as such. For instance, data collection on face-to-face interactions relies on different types of \textsl{sociometric badges} \citep{Eagle2006} for which a recent study reports a false-discovery rate of the event identification of around 20$\%$ when compared to video coded data \citep{Elmer2019}. Political event data on armed conflict, in contrast, are generally collected via automated or human coding of news and social media reporting \citep{Kauffmann2020}. Spurious events may arise in this context if reports of fighting are wrong, as may be the case for propaganda reasons or due to reporters' reliance on rumors, or when fighting took place between different belligerents than those named. Such issues are especially prevalent in machine-coded conflict data where both false-positive and false-discovery rates of over 60$\%$ have been reported \citep{King2003,Jager_2018}. However, even human-coded data suffer from this problem \citep{Dawkins_2020,Weidmann_2015}. 


This discussion suggests that specific types of event data can include unknown quantities of spurious events, which may influence the substantive results obtained from models such as the REM (\citealp{butts2008}) or the Dynamic Actor-Oriented Model \citep{stadtfeld2017,stadtfeld2012}. We thus propose a Relational Events Model with Spurious Events (REMSE) as a method that allows researchers to study relational events from potentially error-prone contexts or data collections methods. Moreover, this tool can assess whether spurious events are observed under a particular model specification and, more importantly, whether they influence the substantive results. The REMSE can thus serve as a straightforward robustness check in situations where the researcher, due to their substantive knowledge, suspects that there are spurious observations and wants to investigate whether they distort their empirical results. 

We take a counting process point of view where some increments of the dyadic counting processes are \textsl{true} events, while others may be attributed to \textsl{spurious} events, i.e., exist due to measurement error. This decomposition results in two different intensities governing the two respective types of events.  The \textsl{spurious} events are described by a \textsl{spurious-event} intensity that we specify independently of the \textsl{true-event} intensity of \textsl{true} events. We present the model under the assumption that the \textsl{spurious} events are purely random. Therefore, we can model the respective intensity solely as a constant term. However, more complex scenarios involving the specification of exogenous and endogenous covariates for the \textsl{spurious-event} intensity are also possible. In general, we are however primarily interested in studying what factors drive the intensity of \textsl{true} events. We model this intensity following \citet{butts2008}, but the methodology is extendable to other model types such as \citet{stadtfeld2017,vu2015, dubois2013_1,perry2013} or \citet{Lerner2021}.

This article is structured as follows: We begin in Section \ref{sec:EcREM} by introducing our methodology. In particular, we lay out the general framework to study relational event data proposed by \citet{butts2008} in Section \ref{sec:REM} and introduce an extension to this framework, the REMSE, to correct for the presence of \textsl{spurious} events in the remainder of Section \ref{sec:EcREM}. Through a simulation study in Section \ref{sec:sim}, we investigate the performance of our proposed estimator when spurious events are correctly specified and when they are nonexistent. We then apply the proposed model in Section \ref{sec:app} to analyze fighting incidents in the Syrian civil war as well as social interaction data from a college campus. A discussion of possible implications and extensions for the analysis of events concludes the article in Section \ref{sec:dis}. 

\section{A Relational Events Model with Spurious Events}
\label{sec:EcREM}
\subsection{Modeling framework for relational events}
\label{sec:REM}

We denote observed events in an event stream $\mathcal{E} = \left\{e_1, ..., e_M \right\}$ of $M$ elements. Each object $e \in \mathcal{E}$ consists of a tuple encoding the information of an event. In particular, we denote the two actors of an event by $a(e)$ and $b(e)$ and the time of the event with $t(e)$. For simplicity of notation, we omit the argument $e$ for $a()$ and $b()$ when no ambiguity exists and write $a_m$ for $a(e_m)$, $b_m$ for $b(e_m)$, and $t_m$ for $t(e_m) ~\forall~ m \in \{1, ..., m\}$. Stemming from our application cases, we mainly focus on undirected events in this article; hence the events $e = (a,b,t)$ and $\tilde{e} = (b,a,t)$ are equivalent in our framework. Note however that the proposed method also generalizes to the directed case. We denote the set of actor-tuples between which events can possibly occur by $\mathcal{R}$, where, for simplicity, we assume that $\mathcal{R}$ is time-constant.

Following \citet{perry2013} and \citet{vu2011}, we assume that the events in $\mathcal{E}$ are generated by an inhomogeneous matrix-valued counting process 
\begin{align}
    \mathbold{N}(t) = (N_{ab}(t)| (a,b) \in \mathcal{R}),
    \label{eq:cp}
\end{align}
which, in our case, is assumed to be a matrix-valued Poisson process (see \citealp{daley2008} for an introduction to stochastic processes). Without loss of generality, we assume that $\mathbold{N}(t)$ is observed during the temporal interval $\mathcal{T}$, starting at $t = 0$. The cells of \eqref{eq:cp} count how often all possible dyadic events have occurred between time $0$ and $t$, hence $\mathbold{N}(t)$ can be conceived as a standard social network adjacency matrix with integer-valued cell entries \citep{Butts2008_b}. For instance, $N_{ab}(t)$ indicates how often actors $a$ and $b$ have interacted in the time interval $[0,t]$. Therefore, observing event $e = (a,b,t)$ constitutes an increase in $N_{ab}(t)$ at time point $t$, i.e. $N_{ab}(t - h)  + 1 =N_{ab}(t)$ for $h \rightarrow 0$.  We denote with $\mathbf{\lambda} (t)$ the matrix-valued intensity of process $\mathbf{N}(t)$. Based on this intensity function we can characterize the instantaneous probability of a unit increase in a specific dimension of $\mathbold{N}(t)$ at time-point $t$ \citep{daley2008}. We parametrize  $\mathbf{\lambda} (t)$ conditional on  the history of the processes, $\mathcal{H}(t)$, which may also include additional exogenous covariates. Hence, $\mathcal{H}(t) = (\mathbold{N}(u),X(u)| u < t )$, where $X(t)$ is some covariate process to be specified later.  Note that we opt for a rather general characterization of Poisson processes, including stochastic intensities that explicitly depend on previous events. We define the intensity function at the tie-level:
\begin{align}
    \lambda_{ab}(t| \mathcal{H}(t), \vartheta) =  \begin{cases}
        \lambda_0(t, \alpha) \exp\{\theta^\top s_{ab}(\mathcal{H}(t))\}, & \text{if } (a,b) \in \mathcal{R} \\
        0, & \text{else}
\end{cases}    
    \label{eq:rem_intensity}
\end{align}
where $\vartheta = (\alpha^\top, \theta^\top)^\top = \text{vec}(\alpha,\theta)$ is defined with the help of a dyadic operator $\text{vec}(\cdot, \cdot)$ that stacks two vectors and $\lambda_0(t,\alpha)$ is the baseline intensity characterized by coefficients $\alpha$, while the parameters $\theta$ weight the statistics computed by $s_{ab}(\mathcal{H}(t))$, which is the function of sufficient statistics. Based on $s_{ab}(\mathcal{H}(t))$, we can formulate endogenous effects, which are calculated from $(N(u)| u < t)$, exogenous variables calculated from $(X(u) | u<t)$, or a combination of the two which results in complex dependencies between the observed events. Examples of endogenous effects for undirected events include degree-related statistics like the absolute difference of the degrees of actors $a$ and $b$ or hyperdyadic effects, e.g., investigating how triadic closure influences the observed events. In our first application case, exogenous factors include a dummy variable whether group $a$ and $b$ share an ethno-religious identity. Alternatively, one may incorporate continuous covariates, e.g., computing the absolute geographic distance between group $a$ and $b$. 
\begin{figure}[t!]
		\centering
		\includegraphics[width=0.8\linewidth,page =1 ]{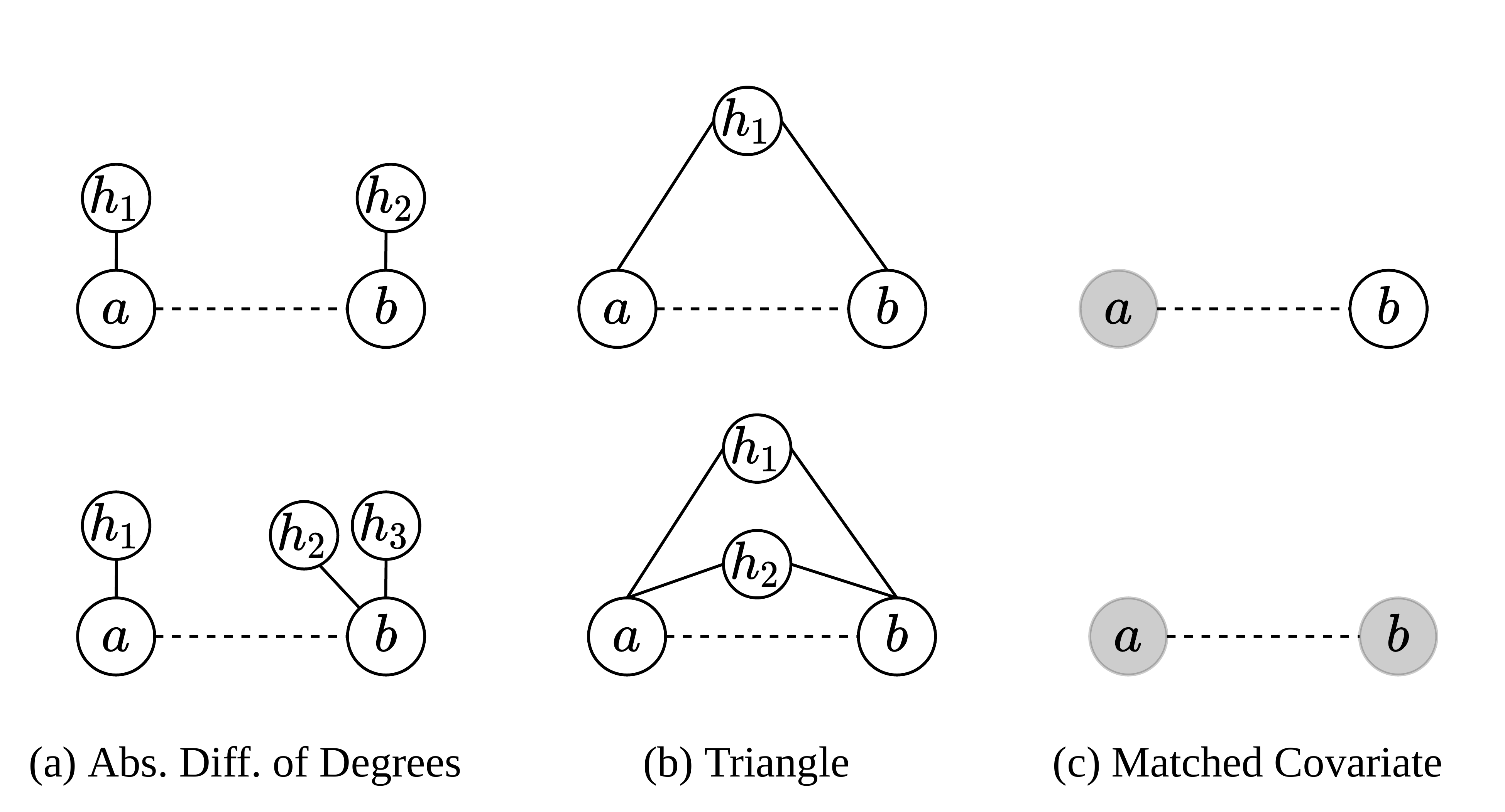}
		\caption{Graphical illustrations of endogenous and exogenous covariates. Solid lines represent past interactions, while dotted lines are possible but unrealized events. Node coloring indicates the node's value on a categorical covariate. The relative risk of the events in the second row compared to the events in the first row is $\exp\{\theta_{end}\}$ if all other covariates are fixed, where $\theta_{end}$ is the coefficient of the respective statistic of each row. 
		}
				\label{fig:effects}
\end{figure}
We give graphical representations of possible endogenous effects in Figure \ref{fig:effects} and provide their mathematical formulations together with a general summary in Annex \ref{sec:annex_stat}. When comparing the structures in the first row with the ones in the second row in Figure \ref{fig:effects}, the respective sufficient statistic of the event indicated by the dotted line differs by one unit. Its intensity thus changes by the multiplicative factor $\exp\{\theta_{endo}\}$, where $\theta_{endo}$ is the respective parameter of the statistic if all other covariates are fixed. The interpretation of the coefficients is, therefore, closely related to the interpretation of relative risk models \citep{Kalbfleisch2002}.    

Previous studies propose multiple options to model the baseline intensity $\lambda_0(t)$. \citet{vu2011,vu2011_2} follow a semiparametric approach akin to the proportional hazard model by \citet{cox1972}, while \citet{butts2008} assumes a constant baseline intensity. We follow \citet{Etezadi-Amoli1987} by setting $\lambda_0(t, \alpha) = \exp\{f(t, \alpha)\}$ , with $f(t, \alpha)$  being a smooth function in time parametrized by B-splines \citep{DeBoor2001}: 
\begin{align}
    f(t, \alpha) = \sum_{k=1}^K \alpha_k B_{k}(t) = \alpha^\top B(t), 
    \label{eq:baseline}
\end{align}
where $B_{k}(t)$ denotes the $k$th B-spline basis function weighted by coefficient $\alpha_k$. To ensure a smooth fit of $f(t, \alpha)$, we impose a penalty (or regularization) on $\alpha$ which is formulated through the priori structure
\begin{align}
p(\alpha) \propto \exp\{- \gamma \alpha^\top \mathbf{S}  \alpha\},    \label{eq:prior_alpha}
\end{align}
 where $\gamma$ is a hyperparameter controlling the level of smoothing and $\mathbf{S}$ is a penalty matrix that penalizes the differences of coefficients corresponding to adjacent basis functions as proposed by \citet{eilers1996}. We ensure identifiability of the smooth baseline intensity by incorporating a sum-to-zero constraint and refer to \citet{Ruppert2003a} and \citet{wood2017} for further details on penalized spline smoothing. Given this notation, we can simplify \eqref{eq:rem_intensity}: 
\begin{align}
    \lambda_{ab}(t| \mathcal{H}(t), \vartheta) =  \begin{cases}
         \exp\{\vartheta^\top \mathcal{X}_{ab}(\mathcal{H}(t), t)\}, & \text{if } (a,b) \in \mathcal{R} \\
        0, & \text{else,} \label{eq:rem_intensity_final}
\end{cases}   
\end{align}
with $\mathcal{X}_{ab}(\mathcal{H}(t),t) = \text{vec}(B(t), s_{ab}(\mathcal{H}(t)))$.
\subsection{Accounting for spurious relational events}

\label{sec:accounting}

Given the discussion in the introduction, we may conclude that some increments of $\mathbf{N}(t)$ are true events, while others stem from spurious events. Spurious events can occur because of coding errors during machine- or human-based data collection. To account for such erroneous data points, we introduce the Relational Events Model with Spurious Events (REMSE). 

\begin{figure}[t!]
		\centering
		\includegraphics[width=0.7\linewidth,page =2 ]{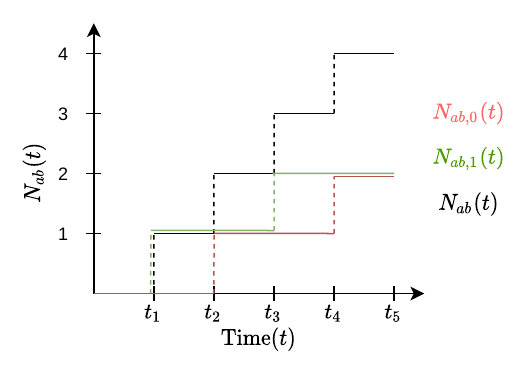}
		\caption{Graphical illustration of a possible path of the counting process of observed events ($N_{ab}(t)$) between actors $a$ and $b$ that encompasses spurious ($N_{ab,0}(t)$) and true events ($N_{ab,1}(t)$).}
				\label{fig:illustration}
\end{figure}

First, we decompose the observed Poisson process into two separate matrix-valued Poisson processes, i.e. $\mathbf{N}(t) = \mathbf{N}_{0}(t)  + \mathbf{N}_1(t) ~ \forall~ t \in \mathcal{T}$. On the dyadic level, $N_{ab,1}(t)$ denotes the number of true events between actors $a$ and $b$ until $t$, and $N_{ab,0}(t)$ the number of events that are spurious. Assuming that $N_{ab}(t)$ is a Poisson process, we can apply the so-called \textsl{thinning} property, stating that two separate processes that sum up to a Poisson process are also Poisson processes \citep{daley2008}. A graphical illustration of the three introduced counting processes, $N_{ab,0}(t), ~N_{ab,1}(t),$ and $N_{ab}(t)$, is given in Figure \ref{fig:illustration}. In this illustrative example, we observe four events at times $t_1, ~t_2, ~t_3,$ and $t_4$, although only the first and third constitute true events, while the second and fourth are spurious. Therefore, the counting process $N_{ab}(t)$ jumps at all times of an event, yet $N_{ab,1}(t)$ does so only at $t_1$ and $t_3$. Conversely, $N_{ab,0}(t)$ increases at $t_2$ and $t_4.$ 

The counting processes $ \mathbf{N}_{0}(t)$ and $\mathbf{N}_1(t)$ are characterized by the dyadic intensities $\lambda_{ab,0}(t|\mathcal{H}_{0}(t), \vartheta_0)$ and $\lambda_{ab,1}(t|\mathcal{H}_{1}(t), \vartheta_1)$, where we respectively denote the history of all spurious and true processes by $\mathcal{H}_{0}(t)$ and $\mathcal{H}_{1}(t)$. This can also be perceived as a competing risks setting, where events can either be caused by the \textsl{true-event} or \textsl{spurious-event} intensity \citep{Alan2000}. To make the estimation of $\theta_0$ and $\theta_1$ feasible and identifiable \citep{Heckman1989}, we assume that both intensities are independent of one another, which means that their correlation is fully accounted for by the covariates. Building on the  \textsl{superpositioning} property of Poisson processes, the specification of those two intensity functions also defines the intensity of the observed counting process $N_{ab}(t)$. In particular, $\lambda_{ab}(t|\mathcal{H}(t), \vartheta)= \lambda_{ab,0}(t|\mathcal{H}_0(t), \vartheta_0) + \lambda_{ab,1}(t|\mathcal{H}_1(t), \vartheta_1)$ holds \citep{daley2008}. 

The \textsl{true-event} intensity $\lambda_{ab,1}(t|\mathcal{H}_{1}(t), \vartheta_1)$ drives the counting process of true events $\mathbf{N}_{1}(t)$ and only depends on the history of true events. This assumption is reasonable since if erroneous events are mixed together with true events, the covariates computed for actors $a$ and $b$ at time $t$ through $s_{ab}(\mathcal{H}(t))$ would be confounded and could not anymore be interpreted in any consistent manner. We specify $\lambda_{ab,1}(t|\mathcal{H}_1(t), \vartheta_1)$ in line with \eqref{eq:rem_intensity} at the dyadic level by: 
\begin{align}
    \lambda_{ab,1}(t| \mathcal{H}_1(t), \vartheta) =  \begin{cases}
         \exp\{\vartheta_1^\top \mathcal{X}_{ab,1}(\mathcal{H}_1(t),t)\}, & \text{if } (a,b) \in \mathcal{R} \\
        0, & \text{else.} \label{eq:design}
\end{cases}   
\end{align}

At the same time, the \textsl{spurious-event} intensity $\lambda_{ab,0}(t|\mathcal{H}_{0}(t), \vartheta_0)$ determines the type of measurement error generating spurious events.  One may consider the spurious-event process as an overall noise level with a constant intensity. This leads to the following setting: 
\begin{align}
   \lambda_{ab,0}(t|\mathcal{H}_{0}(t), \vartheta_0) =  \begin{cases}
        \exp\{\alpha_0\}, & \text{if } (a,b) \in \mathcal{R} \\
        0, & \text{else.}
\end{cases}    
\label{eq:lambda_0}
\end{align}
The error structure, that is, the intensity of the spurious-event process can be made more complex, but to ensure identifiability, $\lambda_{ab,0}(t|\mathcal{H}_{0}(t), \vartheta_0)$ cannot depend on the same covariates as $\lambda_{ab,1}(t| \mathcal{H}_1(t), \vartheta)$. We return to the discussion of this point below and focus on model \eqref{eq:lambda_0} for the moment. 

\subsection{Posterior inference via data augmentation}

To draw inference on $\vartheta = \text{vec}(\vartheta_0, \vartheta_1)$, we employ an empirical Bayes approach. Specifically, we will sample from the posterior of $\vartheta$ given the \textsl{observed} data. Our approach is thereby comparable to the estimation of standard mixture \citep{Diebolt1994} and latent competing risk models \citep{Alan2000}. 


For our proposed method, the \textsl{observed} data is the event stream of all events $\mathcal{E}$ regardless of being a real or a spurious event. To adequately estimate the model formulated in Section \ref{sec:EcREM}, we lack information on whether a given event is spurious or not. We denote this formally as a latent indicator variable $z(e)$ for event $e \in \mathcal{E}$: 
\begin{align*}
z(e) = \begin{cases}
1,& \text{if event $e$ is a true event} \\
0,& \text{if event $e$ is a spurious event}
\end{cases}
\end{align*}
We write $z = (z(e_1), ..., z(e_M))$ to refer to the latent indicators of all events and use $z_m$ to shorten $z(e_m)$. Given this notation, we can apply the data augmentation algorithm developed in \citet{Tanner1987} to sample from the joint posterior distribution of $(Z,\vartheta)$ by iterating between the I Step (Imputation) and P Step (Posterior) defined as: 
\begin{align*}
    &\hspace{-5cm}\text{I Step: Draw $Z^{(d)}$ from the posterior $p(z| \vartheta^{(d-1)}, \mathcal{E})$; } \\
    &\hspace{-5cm}\text{P Step: Draw $\vartheta^{(d)}$ from the augmented $p(\vartheta| z^{(d)}, \mathcal{E})$.}
\end{align*}
This iterative scheme generates a sequence that (under mild conditions) converges to draws from the joint posterior of $(\vartheta,Z)$ and is a particular case of a Gibbs' sampler. Each iteration consists of an Imputation and a Posterior step, resembling the Expectation and Maximization step from the EM algorithm \citep{Dempster1977}. Note, however, that \citet{Tanner1987} proposed this method with multiple imputations in each I Step and a mixture of all imputed \textsl{complete-data} posteriors in the P Step. We follow \citet{Little2002} and \citet{Diebolt1994} by performing one draw of $Z$ and $\vartheta$ in every iteration, which is a specific case of data augmentation. As \citet{Noghrehchi2021} argue, this approach is closely related to the stochastic EM algorithm (\citealp{Celeux1996}). The main difference between the two approaches is that in our P Step, the current parameters are sampled from the \textsl{complete-data} posterior in the data augmentation algorithm and not fixed at its mean as in \citet{Celeux1996}. Consequently, the data augmentation algorithm is a \textsl{proper} multiple imputation procedure (MI, \citealp{rubin1987}), while the stochastic EM algorithm is \textsl{improper} MI (see \citealp{Noghrehchi2021}). We choose the data augmentation algorithm over the stochastic EM algorithm because Rubin's combination rule to get approximate standard errors can only be applied to \textsl{proper} MI procedures \citep{Noghrehchi2021}. 

In what follows, we give details and derivations on the I and P Steps and then exploit MI to combine a relatively small number of draws from the posterior to obtain point and interval estimates for $\vartheta$. 

\paragraph*{Imputation-step:}

To acquire samples from $Z= (Z_1, ..., Z_M)$ conditional on $\mathcal{E}$ and $\vartheta$, we first decompose the joint density by repeatedly applying the Bayes theorem: 
\begin{align}
 \label{eq:iterative}
   p(z |  \vartheta, \mathcal{E}) &= p(z_M, ..., z_1 |  \vartheta, \mathcal{E}) \nonumber \\
   &= \prod_{m= 1}^M p(z_m | z_1, ..., z_{m-1},\vartheta, \mathcal{E}).
\end{align}
The distribution of $z_m$ conditional on $ z_1, ..., z_{m-1},\vartheta$ and $\mathcal{E}$ is: 
\begin{align}
    Z_m | z_1, ..., z_{m-1},\vartheta, \mathcal{E} \sim \text{Bin}\left(1, \frac{\lambda_{a_m b_m,1}(t_{m}|\mathcal{H}_1(t_{m}), \vartheta_1)}{\lambda_{a_m b_m,0}(t_m|\mathcal{H}_0(t_m), \vartheta_0) + \lambda_{a_m b_m,1}(t_m|\mathcal{H}_1(t_m), \vartheta_1)}\right). \label{eq:bin}
\end{align}
Note that the information of $z_1, ... z_{m-1}$ and $\mathcal{E}$ allows us to calculate $ \mathcal{H}_1(t_m)$ as well as $\mathcal{H}_0(t_m)$. 
By iteratively applying \eqref{eq:bin} and plugging in $\vartheta^{(d)}$ for $\vartheta$, we can draw samples in the I Step of $Z= (Z_1, ..., Z_M)$ through a sequential design that sweeps once from  $Z_1$ to $Z_M$. The mathematical derivation of \eqref{eq:bin} is provided in Annex \ref{sec:appenix_da}.

\paragraph*{Posterior-step:}

As already stated, we assume that the \textsl{true-event} and \textsl{spurious-event} intensities are independent. Hence, the sampling from the \textsl{complete-data} posteriors of $\vartheta_0$ and $\vartheta_1$ can be carried out independently. In the ensuing section, we therefore only show how to sample from $\vartheta_1|  z, \mathcal{E}$, but sampling from $\vartheta_0| z, \mathcal{E}$ is possible in the same manner. To derive this posterior, we begin by showing that the likelihood of $\mathcal{E}$ and $z$ with parameter $\vartheta_1$ is the likelihood of the counting process $\mathbf{N}_1(t)$, which resembles a Poisson regression. Consecutively, we state all priors to derive the desired \textsl{complete-data} posterior. 


Given a general $z$ sampled in the previous I Step and $\mathcal{E}$, we reconstruct a unique complete path of $\mathbf{N}_1(t)$  by setting 
\begin{align}
    N_{ab,1}(t) = \sum_{\substack{ e \in \mathcal{E}; \\  z(e) = 1, ~ t(e)\leq t} }  \mathbb{I}(a(e) =a, b(e) = b) ~ \forall~ (a,b) \in \mathcal{R},~ t\in \mathcal{T}, \label{eq:reconstruct}
\end{align}
where $\mathbb{I}(\cdot)$ is an indicator function. The corresponding likelihood of $\mathbf{N}_1(t)$ results from the property that any element-wise increments of the counting process between any times $s$ and $t$ with $t>s$ and arbitrary actors $a$ and $b$ with $(a,b) \in \mathcal{R}$ are Poisson distributed: 
\begin{align}
      N_{ab,1}(t) - N_{ab,1}(s) \sim \text{Pois}\left(\int_s^t \lambda_{ab,1}\left( u|\mathcal{H}_1(u),\vartheta_1\right) du\right). \label{eq:pp_norm}
\end{align}
The integral in \eqref{eq:pp_norm} is approximated through simple rectangular approximation between the observed event times to keep the numerical effort feasible, so that the distributional assumption simplifies to: 
\begin{align}
      Y_{ab,1}(t_m) = N_{ab,1}(t_m) - N_{ab,1}(t_{m-1}) \sim& \text{Pois}\left(\left(t_m - t_{m-1}\right) \lambda_{ab,1}\left(t_m|\mathcal{H}_1\left(t_m\right),\vartheta_1\right)\right) \label{eq:pp} \\
      ~\forall ~ m \in \{1, ..., M\} &\text{ with }z_m = 1\text{ and }(a,b) \in \mathcal{R}. \nonumber
\end{align}



We specify the priors for $\alpha_1$ and $\theta_1$ separately and independent of one another. The prior for $\alpha_1$ was already stated in \eqref{eq:prior_alpha}. Through a restricted maximum likelihood approach, we estimate the corresponding hyperparameter $\gamma_1$ such that it maximizes the marginal likelihood of  $z$ and $\mathcal{E}$ given $\gamma_1$ (for additional information on this estimation procedure and general empirical Bayes theory for penalized splines see \citealp{Wood2011b,Wood2020}). Regarding the linear coefficients $\theta_1$, we assume flat priors, i.e. $p(\theta_1) \propto k$, indicating no prior knowledge. 
 
In the last step, we apply Wood's (\citeyear{Wood2006}) result that for large samples, the posterior distribution of $\vartheta_1$ under likelihoods resulting from distributions belonging to the exponential family, such as the Poisson distribution in \eqref{eq:pp}, can be approximated through: 
\begin{align}
    \vartheta_1 | z, \mathcal{E}  \sim N\left(\hat{\vartheta}_1, \mathbf{V}_1\right). \label{eq:aug_posterior}
\end{align}
 Here, $\hat{\vartheta}_1$ denotes the penalized maximum likelihood estimator resulting from \eqref{eq:pp} with the extended penalty matrix $\tilde{ \mathbf{S}}_1$ defined by 
 \begin{align*}
    \tilde{ \mathbf{S}}_1 = \begin{bmatrix}
 \mathbf{S}_1 & \mathbf{O}_{p \times q} \\
\mathbf{O}_{p \times q} & \mathbf{O}_{q \times q}
\end{bmatrix}
\end{align*} 
 with $\mathbf{O}_{p \times q} \in \mathbb{R}^{p\times q}$ for $p,q \in \mathbb{N}$ being a matrix filled with zeroes and $\mathbf{S}_1$ defined in accordance with \eqref{eq:prior_alpha}. For $\vartheta_1 = \text{vec}(\alpha_1, \theta_1),$ let $p$ be the length of $\alpha_1$ and $q$ of $\theta_1$. The penalized likelihood is then given by: 
\begin{align}
    \ell_p(\vartheta_1; z,\mathcal{E})  = \ell(\vartheta_1; z,\mathcal{E}) - \gamma_1\vartheta_1^\top   \tilde{ \mathbf{S}}_1   \vartheta_1 \label{eq:penph},
\end{align}
which is equivalent to a generalized additive model; hence we refer to \citet{wood2017} for a thorough treatment of the computational methods needed to find $\hat{\vartheta}_1$. 
The variance matrix in \eqref{eq:aug_posterior} has the following structure: 
\begin{align*}
    \mathbf{V}_1 = \left(\mathcal{X}_1^\top \mathbf{W}_1 \mathcal{X}_1 + \gamma_1 \tilde{ \mathbf{S}}_1  \right)^{-1}.
\end{align*}
Values for $\gamma_1$ and $\hat{\vartheta}_1$ can be extracted from the estimation procedure to maximize \eqref{eq:penph} with respect to $\vartheta_1$, while $\mathcal{X}_1 \in \mathbb{R}^{(M |\mathcal{R}|)\times (p + q)}$ is a matrix whose rows are given by $\mathcal{X}_{ab,1}(\mathcal{H}_1(t_m),t_{m-1})$ as defined in \eqref{eq:design} for $m \in \{ 1, ..., M\}$ and $(a,b) \in \mathcal{R}$. Similarly,  \newline$\mathbf{W}_1 = \text{diag}\big(\lambda_{ab,1}(t| \mathcal{H}_1(t), \vartheta_1);t \in \{ t_1, ..., t_M\}, (a,b) \in \mathcal{R}\big)$ is a diagonal matrix. 


\begin{algorithm}[t!]
	\SetAlgoLined
	\KwResult{$(\vartheta^{(1)},z^{(1)}), ..., (\vartheta^{(D)},z^{(D)})$}
	\emph{Set:} $\vartheta^{(0)}$ to be the posterior mean of the \textsl{true} and \textsl{spurious} events, which are sampled randomly from the observed events with equal probability \\
	\For{$d \in \{1, ..., D\}$}{
	    \textbf{Imputation Step: Sample $Z^{(d)}| \vartheta^{(d-1)}, \mathcal{E}$} \\
		\For{$m \in \lbrace 1, ..., M\rbrace$}{
			\begin{itemize}
			    \item Sample $Z^{(d)}_m | z^{(d)}_1,..., z^{(d)}_{m-1}, \vartheta^{(d-1)}, \mathcal{E}$ according to \eqref{eq:bin}
			    \item  If $z^{(d)}_m = 1$  update $s_{ab}(\mathcal{H}_1(t_m)) ~ \forall ~ (a,b) \in \mathcal{R}$
			\end{itemize}
		} 
		 \textbf{Posterior Step: Sample $\vartheta^{(d)}| z^{(d)}, \mathcal{E}$} \\
		 \begin{itemize}
		     \item Reconstruct $\mathbf{N}_0(t)$ and $\mathbf{N}_1(t) ~\forall~ t\in \mathcal{T}$ from $ z^{(d)}$ and $\mathcal{E}$ according to \eqref{eq:reconstruct}
		     \item Obtain $\hat{\vartheta}_0$ and $\mathbf{V_0}$ by maximizing the penalized Poisson likelihood \newline given in \eqref{eq:pp} (only for  $\mathbf{N}_0(t)$ instead of  $\mathbf{N}_1(t)$)
		     \item Sample $\vartheta_0^{(d)}| z^{(d)}, \mathcal{E} \sim N(\hat{\vartheta}_0, \mathbf{V_0})$
		     \item Obtain $\hat{\vartheta}_1$ and $\mathbf{V_1}$ by maximizing the penalized Poisson likelihood \newline given in \eqref{eq:pp}
		     \item Sample $\vartheta_1^{(d)}| z^{(d)}, \mathcal{E} \sim N(\hat{\vartheta}_1, \mathbf{V_1})$
		 \end{itemize}
	}
	\caption{Pseudo-Code to obtain $D$ samples from the data augmentation algorithm.}
	\label{al:da_pseudo}
\end{algorithm}

For the P Step, we now plug in $z^{(d)}$ for $z$ in \eqref{eq:aug_posterior} to obtain $\hat{\vartheta}_1$ and $\mathbf{V}_1$ by carrying out the corresponding \textsl{complete-case} analysis. In the case where  no spurious events exist, the complete estimation can be carried out in a single P Step. In Algorithm \ref{al:da_pseudo}, we summarize how to generate a sequence of random variables according to the data augmentation algorithm. 

\paragraph*{Multiple imputation:}

One could use the data augmentation algorithm to get a large amount of samples from the joint posterior of $(\vartheta, Z)$ to calculate empirical percentiles for obtaining any types of interval estimates. However, in our case this endeavor would be very time-consuming and even infeasible. To circumvent this, \citet{Rubin1976} proposed multiple imputation as a method to approximate the posterior mean and variance. Coincidentally, the method is especially successful when the \textsl{complete-data} posterior is multivariate normal as is the case in \eqref{eq:aug_posterior}, thus only a small number of draws is needed to obtain good approximations \citep{Little2002}. To be specific, we apply the law of iterative expectation and variance: 
\begin{align}
    \mathbb{E}(\vartheta|\mathcal{E}) &= \mathbb{E}(\mathbb{E}(\vartheta|\mathcal{E}, z)|z) \label{eq:int_ew}\\
    \text{Var}(\vartheta|\mathcal{E})  &=  \mathbb{E}(\text{Var}(\vartheta|\mathcal{E}, z)|z) +\text{Var}(\mathbb{E}(\vartheta|\mathcal{E},z)|z) \label{eq:int_var}.
\end{align}
Next, we approximate \eqref{eq:int_ew} and \eqref{eq:int_var} using a Monte Carlo quadrature with $K$ samples from the posterior obtained via the data augmentation scheme summarized in Algorithm \ref{al:da_pseudo} after a burn-in period of $D$ iterations: 
\begin{align}
    \mathbb{E}(\vartheta|\mathcal{E}_{\text{obs}}) &\approx \frac{1}{K} \sum_{k = D + 1}^{D+K} \hat{\vartheta}^{(k)} = \bar{\vartheta} \label{eq:post_mean} \\
    \text{Var}(\vartheta|\mathcal{E}_{\text{obs}})  &\approx \frac{1}{K} \sum_{k = D + 1}^{D+K} \mathbf{V}^{(k)} + \frac{K+1}{K(K-1)}\sum_{k = D + 1}^{D+K} \left(\hat{\vartheta}^{(k)} - \bar{\vartheta}\right) \left(\hat{\vartheta}^{(k)} - \bar{\vartheta}\right)^\top\nonumber \\
    &= \bar{\mathbf{V}} + \bar{\mathbf{B}},\label{eq:post_var} 
\end{align}
where $\hat{\vartheta}^{(k)} = \text{vec}\left(\hat{\vartheta}^{(k)}_0, \hat{\vartheta}^{(k)}_1\right)$ encompasses the \textsl{complete-data} posterior means from the $k$th sample and $\mathbf{V}^{(k)} = \text{diag} \big( \mathbf{V}^{(k)}_0 ,  \mathbf{V}^{(k)}_1\big)$ is composed of the corresponding variances defined in \eqref{eq:aug_posterior}.
We can thus construct point and interval estimates from relatively few draws of the posterior based on a multivariate normal reference distribution \citep{Little2002}. 



\section{Simulation Study}
\label{sec:sim}

We conduct a simulation study to explore the performance of the REMSE compared to a REM, which assumes no spurious events, in two different scenarios, including a regime where measurement error is correctly specified in the REMSE and one where spurious events are instead non-existent. 

\paragraph*{Simulation design:} In $S=1000$ runs, we simulate event data between $n = 40$ actors under known true and spurious intensity functions in each example. For exogenous covariates, we generate categorical and continuous actor-specific covariates, transformed to the dyad level by checking for equivalence in the categorical case and computing the absolute difference for the continuous information. Generally, we simulate both counting processes $\mathbf{N_1(t)}$ and $\mathbf{N_0(t)}$ separately and stop once $|\mathcal{E}_1| = 500$. 

The data generating processes for \textsl{true} events is identical in each case and given by:
\begin{align*}
    \lambda_{ab,1}(t|\mathcal{H}_1(t), \vartheta_1) &= \exp \{-5 + 0.2\cdot s_{ab,Degree~Abs.}(\mathcal{H}_1(t))  \tag{DG 1-2} \\
    & \hspace{1.4cm}  +  0.1\cdot s_{ab, Triangle}(\mathcal{H}_1(t)) - 0.5\cdot s_{ab, Repetition ~ Count}(\mathcal{H}_1(t)) \\
    & \hspace{1.4cm} + 2\cdot s_{ab, Sum ~ cont.}(\mathcal{H}_1(t))  - 2\cdot s_{ab, Match ~ cat.}(\mathcal{H}_1(t))   \}, \nonumber
\end{align*} 
where we draw the continuous exogenous covariate (cont.) from a standard Gaussian distribution and the categorical exogenous covariates (cat.) from a categorical random variable with seven possible outcomes, all with the same probability. Mathematical definition of the endogenous and exogenous statistics are given in Annex \ref{sec:annex_stat}. In contrast, the \textsl{spurious-event}  intensity differs across regimes to result in correctly specified \eqref{eq:dg1} and nonexistent \eqref{eq:dg2} measurement errors: 
\begin{align*}
    \lambda_{ab,0}(t| \mathcal{H}_0(t),\vartheta_0) &= \exp \{-2.5 \}    \tag{DG 1} \label{eq:dg1}\\
    \lambda_{ab,0}(t| \mathcal{H}_0(t),\vartheta_0) &= 0  \tag{DG 2} \label{eq:dg2}
\end{align*}
Given these intensities, we follow \citet{dubois2013_1} to sample the events. 

\begin{table}[!tbp]
\caption{Result of the simulation study for the REMSE and REM 
      with the two data-generating processes (DG 1, DG 2). 
      For each DG and covariate, we note the AVE (AVerage Estimate), RMSE (Root-Mean-Squared Error), 
      and CP (Coverage Probability). We report the average Percentage of False Events (PFE) for each DG in the last row. 
      \label{tbl:simulation_res}} 
\begin{center}
\begin{tabular}{lrcrrrcrrr}
\hline
\multicolumn{1}{l}{\  ~}&\multicolumn{1}{c}{\  }&\multicolumn{1}{c}{\  }&\multicolumn{3}{c}{\  REMSE}&\multicolumn{1}{c}{\  }&\multicolumn{3}{c}{\  REM}\tabularnewline
\cline{4-6} \cline{8-10}
\multicolumn{1}{l}{}&\multicolumn{1}{c}{Coefs.}&\multicolumn{1}{c}{}&\multicolumn{1}{c}{AVE}&\multicolumn{1}{c}{RMSE}&\multicolumn{1}{c}{CP}&\multicolumn{1}{c}{}&\multicolumn{1}{c}{AVE}&\multicolumn{1}{c}{RMSE}&\multicolumn{1}{c}{CP}\tabularnewline
\hline
{\  DG 1 (PFE: 4.819 $\%$)}&&&&&&&&&\tabularnewline
~~Intercept&$-5.0$&&$-4.936$&$0.337$&$0.944$&&$-3.510$&$1.523$&$0.003$\tabularnewline
~~Degree abs&$ 0.2$&&$ 0.198$&$0.009$&$0.940$&&$ 0.168$&$0.033$&$0.018$\tabularnewline
~~Triangle&$ 0.1$&&$ 0.101$&$0.019$&$0.949$&&$ 0.094$&$0.019$&$0.932$\tabularnewline
~~Repetition&$-0.5$&&$-0.494$&$0.035$&$0.946$&&$-0.385$&$0.120$&$0.039$\tabularnewline
~~Cov. cont.&$ 2.0$&&$ 1.982$&$0.101$&$0.951$&&$ 1.557$&$0.453$&$0.003$\tabularnewline
~~Cov. cat.&$-2.0$&&$-1.986$&$0.246$&$0.952$&&$-1.594$&$0.461$&$0.515$\tabularnewline
~~$\widehat{PFE}$ (in $\%$)& &&&$4.835$&&&&&\tabularnewline\hline
{\  DG 2 (PFE 0 $\%$)}&&&&&&&&&\tabularnewline
~~Intercept&$-5.0$&&$-5.040$&$0.286$&$0.954$&&$-5.027$&$0.281$&$0.955$\tabularnewline
~~Degree abs&$ 0.2$&&$ 0.201$&$0.008$&$0.958$&&$ 0.201$&$0.008$&$0.956$\tabularnewline
~~Triangle&$ 0.1$&&$ 0.102$&$0.018$&$0.955$&&$ 0.102$&$0.018$&$0.952$\tabularnewline
~~Repetition&$-0.5$&&$-0.505$&$0.030$&$0.969$&&$-0.504$&$0.030$&$0.964$\tabularnewline
~~Cov. cont.&$ 2.0$&&$ 2.009$&$0.087$&$0.952$&&$ 2.006$&$0.086$&$0.948$\tabularnewline
~~Cov. cat.&$-2.0$&&$-2.007$&$0.231$&$0.952$&&$-2.004$&$0.230$&$0.952$\tabularnewline
~~$\widehat{PFE}$ (in $\%$)& &&&$0.0001$&&&&&\tabularnewline
\hline
\end{tabular}\end{center}
\end{table}

Although the method is estimated in a Bayesian framework, we can still assess the frequentist properties of the estimates of the REMSE and REM.  In particular, the average point estimate (AVE), the root-mean-squared error (RMSE) and the coverage probabilities (CP) are presented in Table \ref{tbl:simulation_res}. The AVE of a specific coefficient is the average over the posterior modes in each run: 
\begin{align*}
    \text{AVE} = \frac{1}{S} \sum_{s = 1}^S \bar{\vartheta}_s,
\end{align*}
where $\bar{\vartheta}_t$ is the posterior mean \eqref{eq:post_mean} of the $t$th simulation run. To check for the average variance of the error in each run, we further report the RMSEs of estimating the coefficient vector $\vartheta$: 
\begin{align*}
    \text{RMSE} =  \sqrt{\frac{1}{S} \sum_{s = 1}^S\left(\bar{\vartheta}_s - \vartheta \right)^\top \left(\bar{\vartheta}_s - \vartheta\right)},
\end{align*}
where $\vartheta$ is the ground truth coefficient vector defined above. Finally, we assess the adequacy of the uncertainty quantification by computing the percentage of runs in which the real parameter lies within the confidence intervals based on a multivariate normal posterior with mean and variance given in \eqref{eq:post_mean} and \eqref{eq:post_var}. According to standard statistical theory for interval estimates, this coverage probability should be around $95\%$ \citep{casella2001}.

 \paragraph*{Results:}\ref{eq:dg1} shows how the estimators behave if the true and false intensities are correctly specified. The results in Table \ref{tbl:simulation_res} suggest that the REMSE can recover the coefficients from the simulation. On the other hand, strongly biased estimates are obtained in the REM, where not only the average estimates are biased, but we also observe high RMSEs and violated coverage probabilities. 

In the second simulation, we assess the performance of the spurious event model when it is misspecified. In particular, we investigate what happens when there are no spurious events in the data, i.e., all events are real, and the intensity of $N_{ab,2}(t)$ is zero in \ref{eq:dg2}.  Unsurprisingly, the REM allows for valid and unbiased inference under this regime. But our stochastic estimation algorithm proves to be robust as for most runs, the simulated events were at some point only consisting of true events. In other words, the REMSE can detect the spurious events correctly and is unbiased if none occur in the observed data. 

For both \ref{eq:dg1} and \ref{eq:dg2}, the PFE estimated by the REMSE closely matches the observed one whereas the REM, by constraining it to zero, severely underestimates the PFE in \ref{eq:dg1}. In sum, the simulation study thus offers evidence that the REMSE increases our ability to model relational event data in the presence of measurement error 
while being equivalent to a standard REM when spurious events do not exist in the data.

\section{Application}
\label{sec:app}

\begin{table}[t]
    \centering
    \caption{Descriptive information on the two analyzed data sets.}
    \begin{tabular}{l c c } \hline
         & Conflict Event Data in \ref{sec:app_syria} & Co-location Events in \ref{sec:app_housing}\\ \hline
        Source &  ACLED & MIT Human Dynamics Lab \\
         &  \citep{Raleigh_Linke_Hegre_Karlsen_2010} & \citep{Madan_Cebrian_Moturu_Farrahi_Pentland_2012}\\
        Observational Period & 2017:01:01 - 2019:01:01  &  2008:11:01 - 2008:11:04 \\
        Number of Actors & 68\footnote{ We include all actors that, within the two-year period, participated in at least five events. To verify their existence and obtain relevant covariates, we compared them first to data collected by \citet{Gade_Hafez_Gabbay_2019} and then to various sources including news media reporting. We omitted two actors on which we could not find any information as well as actor aggregations such as ``rioters'' or ``syrian rebels''.}   & 58\\
        Number of Events &  4,362   & 2,489\footnote{To capture new events instead of repeated observations of the same event, we omit events where the most recent previous interaction between \emph{a} and \emph{b} occurred less than 20 minutes before.}\\
    \hline    
    \end{tabular}
    \label{tab:appl_info}
\end{table}

Next we apply the REMSE on two real-world data sets motivated by the types of event data discussed in the introduction, namely human-coded conflict events in the Syrian civil war and co-location event data generated from the Bluetooth devices of students in a university dorm. Information on the data sources, observational periods and numbers of actors and events is summarized in Table \ref{tab:appl_info}. 
Following the above presentation, we focus on modeling the \textsl{true-event} intensity of the REMSE and limit the  \textsl{spurious-event} intensity to the constant term. Covariates are thus only specified for the \textsl{true-event} intensity. In our applications, the samples drawn according to Algorithm \ref{al:da_pseudo} converged to a stationary distribution within the first 30 iterations. To obtain the reported point and interval estimates via MI, we sampled 30 additional draws. Due to space restrictions, we keep our discussions of the substantive background and results of both applications comparatively short.

\subsection{Conflict events in the Syrian civil war}
\label{sec:app_syria}

In the first example, we model conflict events between different belligerents as driven by both exogenous covariates and endogenous network mechanisms. The exogenous covariates are selected based on the literature on inter-rebel conflict. We thus include dummy variables indicating whether two actors share a common ethno-religious identity or receive material support by the same external sponsor as these factors have previously been found to reduce the risk of conflict \citep{Popovic_2018, Gade_Hafez_Gabbay_2019}. Additionally, we include binary indicators of two actors being both state forces or both rebel groups as conflict may be less likely in the former but more likely in the latter case \citep{Dorff_Gallop_Minhas_2020}. 

Furthermore, we model endogenous processes in the formation of the conflict event network and consider four statistics for this purpose. First, we account for repeated fighting between two actors by including both the count of their previous interactions as well as a binary indicator of repetition, which takes the value 1 if that count is at least 1. We use this additional endogenous covariate as a conflict onset arguably comprises much more information than subsequent fighting. Second, we include the absolute difference in \emph{a} and \emph{b}'s degree to capture whether actors with a high extent of previous activity are prone to engage each other or, instead, tend to fight less established groups to pre-empt their rise to power. Finally, we model hyper-dyadic dependencies by including a triangle statistic that captures the combat network's tendency towards triadic closure. 

Given that fighting should be a relatively obvious event, one may wonder why conflict event data may include spurious observations. This is because all common data collection efforts on armed conflict cannot rely on direct observation but instead use news and social media reporting. Spurious events thus occur when these sources report fighting which did not actually take place as such. In armed conflict, this can happen for multiple reasons. For instance, pro-government media may falsely report that state security forces engaged with and defeated rebel combatants to boost morale and convince audiences that the government is winning. Social media channels aligned with a specific rebel faction may similarly claim victories by its own forces or, less obviously, battles where a rival faction fought and suffered defeat against another group. In war-time settings, journalists may also be unable or unwilling to enter conflict areas and thus base their reporting on local contacts, rumors, or hear-say. Finally, spurious observations may arise here when reported fighting occurred but was attributed to the wrong belligerent faction at some point in the data collection process. 
From a substantive perspective, it is thus advisable to check for the influence of spurious events when analysing these data. 

\begin{table}[!tbp]
\caption{Combat events in the Syrian civil war: Estimated coefficients with confidence intervals noted in brackets in the first column,
      while the Z values are given in the second column. 
      The results of the REMSE are given in the first two columns, 
      while the coefficients of the REM are depicted in the last two columns. The last row reports the estimated average Percentage of False Events (PFE). \label{tbl:res_syria}} 
\begin{center}
\begin{tabular}{lccccc}
\hline
\multicolumn{1}{l}{\  }&\multicolumn{2}{c}{\  REMSE}&\multicolumn{1}{c}{\  }&\multicolumn{2}{c}{\  REM}\tabularnewline
\cline{2-3} \cline{5-6}
\multicolumn{1}{l}{}&\multicolumn{1}{c}{Coef./CI}&\multicolumn{1}{c}{Z Val.}&\multicolumn{1}{c}{}&\multicolumn{1}{c}{Coef./CI}&\multicolumn{1}{c}{Z Val.}\tabularnewline
\hline
Intercept&-10.047&-102.124&&-9.944&-115.723\tabularnewline
&[-10.24,-9.854]&&&[-10.112,-9.775]&\tabularnewline
Degree Abs&0.03&12.677&&0.03&17.295\tabularnewline
&[0.026,0.035]&&&[0.027,0.034]&\tabularnewline
Repetition Count &0.009&45.28&&0.009&56.342\tabularnewline
&[0.009,0.01]&&&[0.009,0.01]&\tabularnewline
First Repetition&5.052&54.321&&4.911&64.946\tabularnewline
&[4.87,5.235]&&&[4.763,5.059]&\tabularnewline
Triangle&0.074&10.229&&0.073&18.989\tabularnewline
&[0.06,0.089]&&&[0.065,0.08]&\tabularnewline
Match Ethno-Religious Id.&-0.387&-5.225&&-0.393&-5.852\tabularnewline
&[-0.532,-0.242]&&&[-0.525,-0.262]&\tabularnewline
Match Rebel&0.159&3.27&&0.171&4.381\tabularnewline
&[0.064,0.255]&&&[0.094,0.247]&\tabularnewline
Match State Force&-0.087&-0.721&&-0.077&-0.723\tabularnewline
&[-0.323,0.149]&&&[-0.287,0.132]&\tabularnewline
Common Sponsor 1&-0.218&-2.51&&-0.227&-2.957\tabularnewline
&[-0.388,-0.048]&&&[-0.378,-0.077]&\tabularnewline
$\widehat{PFE}$ (in $\%$) & 1.1& && 0& \tabularnewline
\hline
\end{tabular}\end{center}
\end{table}

Table \ref{tbl:res_syria} accordingly presents the results of an REM and the  REMSE. Beginning with the exogenous covariates, belligerents are found to be less likely to fight each other when they share an ethno-religious identity or receive resources from the same external sponsor. In contrast, there is no support for the idea that state forces exhibit less fighting among each other than against rebels in this type of internationalized civil war, whereas different rebel groups are more likely to engage in combat against one another. Furthermore, we find evidence that endogenous processes affect conflict event incidence. The binary repetition indicator exhibits the strongest effect across all covariates, implying that two actors are more likely to fight each other if they have done so in the past. As indicated by the positive coefficient of the repetition count, the dyadic intensity further increases the more they have previously fought with one another. The absolute degree difference also exhibits a positive effect, meaning that fighting is more likely between groups with different levels of previous activity. And finally, the triangle statistic's positive coefficient suggests that even in a fighting network, triadic closure exists. This may suggest that belligerents engage in multilateral conflict, attacking the enemy of their enemy, in order to preserve the existing balance of capabilities or change it in their favor \citep{Pischedda_2018}.

This discussion holds for both the results of REM and REMSE. Their point estimates are generally quite similar in this application, suggesting that spurious events do not substantively affect empirical results in this case. That being said, there are two noticeable differences between the two models. First, the coefficient estimates for the binary indicator of belligerents having fought before differs between the two models. In the REM, it implies a multiplicative change of $\exp\{4.911\}=135.775$ while for the REMSE, it is estimated at $\exp\{5.059\}=157.433$. While both models thus identify this effect to be positive and significant, it is found to be substantively stronger when spurious events are accounted for. Second, the two models differ in how precise they deem estimates to be. This difference is clearest in their respective Z-values, which are always farther away from zero for the REM than the REMSE. 
As a whole, these results nonetheless show that spurious events have an overall small influence on substantive results in this application. The samples from the latent indicators $z$ also indicate that only approximately 1$\%$ of the observations, about 50 events, are on average classified as spurious events. These findings offer reassurance for the increasing use of event data to study armed conflict.

\subsection{Co-location events in university housing}
\label{sec:app_housing}

In our second application, we use a subset of the co-location data collected by \citet{Madan_Cebrian_Moturu_Farrahi_Pentland_2012} to model when students within an American university dorm interact with each other. These interactions are deduced from continuous (every 6 minutes) scans of proximity via the Bluetooth signals of students' mobile phones. \citet{Madan_Cebrian_Moturu_Farrahi_Pentland_2012} used questionnaires to collect a host of information from the participating students. This information allows us to account for both structural and more personal exogenous predictors of social interaction. We thus include binary indicators of whether two students are in the same year of college or live on the same floor of the dorm to account for the expected homophily of social interactions \citep{McPherson2001}. In addition, we incorporate whether two actors consider each other close friends\footnote{We symmetrized the friendship network, i.e., if student $a$ nominated student $b$ as a close friend, we assume that the relationship is reciprocated.}. Given that the data were collected around a highly salient political event, the 2008 US presidential election, we also incorporate a dummy variable to measure whether they share the same presidential preference and a variable measuring their similarity in terms of interest in politics \citep{Butters_Hare_2020}. In addition, we include the same endogenous network statistics here as in section \ref{sec:app_syria}. These covariates allow us to capture the intuitions that individuals tend to socialize with people that they have interacted with before, are not equally popular as they are, and they share more common friends with \citep{rivera2010}. Compared to the first application, sources of spurious events here are more evident as students may not actually interact with but be physically close to and even face each other, e.g., riding an elevator, queuing in a store, or studying in a common space.   

\begin{table}[!tbp]
\caption{Co-location Events in University Housing: Estimated coefficients with confidence intervals noted in brackets in the first column,
      while the Z values are given in the second column. 
      The results of the REMSE are given in the first two columns, 
      while the coefficients of the REM are depicted in the last two columns.  The last row reports the estimated average Percentage of False Events (PFE). \label{tbl:res_sensor}} 
\begin{center}
\begin{tabular}{lccccc}
\hline
\multicolumn{1}{l}{\  }&\multicolumn{2}{c}{\  REMSE}&\multicolumn{1}{c}{\  }&\multicolumn{2}{c}{\  REM}\tabularnewline
\cline{2-3} \cline{5-6}
\multicolumn{1}{l}{}&\multicolumn{1}{c}{Coef./CI}&\multicolumn{1}{c}{Z Val.}&\multicolumn{1}{c}{}&\multicolumn{1}{c}{Coef./CI}&\multicolumn{1}{c}{Z Val.}\tabularnewline
\hline
 Intercept &-10.077&-124.905&&-10.012&-139.269\tabularnewline
&[-10.235,-9.919]&&&[-10.153,-9.871]&\tabularnewline
Degree Abs&0.025&5.361&&0.025&6.369\tabularnewline
&[0.016,0.035]&&&[0.017,0.032]&\tabularnewline
Repetition Count&0.066&27.263&&0.065&29.988\tabularnewline
&[0.061,0.07]&&&[0.061,0.069]&\tabularnewline
First Repetition&2.714&42.024&&2.615&44.704\tabularnewline
&[2.587,2.84]&&&[2.501,2.73]&\tabularnewline
Triangle&0.049&6.597&&0.049&8.109\tabularnewline
&[0.035,0.064]&&&[0.037,0.061]&\tabularnewline
Match Floor&0.117&2.197&&0.123&2.439\tabularnewline
&[0.013,0.221]&&&[0.024,0.222]&\tabularnewline
Match Presidential Pref&0.195&4.374&&0.188&4.499\tabularnewline
&[0.108,0.282]&&&[0.106,0.27]&\tabularnewline
Match Year&-0.003&-0.051&&-0.012&-0.236\tabularnewline
&[-0.109,0.104]&&&[-0.112,0.088]&\tabularnewline
Dyad. Friendship&0.157&3.145&&0.15&3.15\tabularnewline
&[0.059,0.254]&&&[0.057,0.243]&\tabularnewline
Sim. Interested In Politics&-0.018&-0.74&&-0.021&-0.917\tabularnewline
&[-0.064,0.029]&&&[-0.065,0.024]&\tabularnewline
$\widehat{PFE}$ (in $\%$) & 3.264 & && 0& \tabularnewline
\hline
\end{tabular}\end{center}
\end{table}

We present the results in Table \ref{tbl:res_sensor}. Beginning with the exogenous covariates, we find that the observed interactions tend to be homophilous in that students have social encounters with people they live together with, consider their friends, and share a political opinion with. In contrast, neither a common year of college nor a similar level of political interest are found to have a statistically significant effect on student interactions. At the same time, these results indicate that the social encounters are affected by endogenous processes. Having already had a previous true event is found to be the main driver of the corresponding intensity; hence having a very strong and positive effect. Individuals who have socialized before are thus more likely to socialize again, an effect that, as indicated by the repetition count, increases with the number of previous interactions. Turning to the other endogenous covariates, the result for absolute degree difference suggests that students $a$ and $b$ are more likely to engage with each other if they have more different levels of previous activity, suggesting that e.g. popular individuals attract attention from less popular ones. As is usual for most social networks \citep{newman2003}, the triangle statistic is positive, meaning that students ``\textsl{socialize}'' with the friends of their friends. 

As in the first application, the REM and REMSE results presented in Table \ref{tbl:res_sensor} are closely comparable but also show some differences. Again, the effect estimate for binary repetition, at $\exp\{2.715\}=15.105$, is higher in the REMSE than in the REM ($\exp\{2.615\}=13.667$) while Z-values and confidence intervals obtained in the REM are substantially smaller in the REM than in the REMSE. In the co-location data too, the results are thus not driven by the presence of spurious events but accounting for these observations does affect results to some, albeit rather negligible, extent. This is the case even though
the average percentage of spurious events here is comparatively high at 3$\%$. That leaving out the corresponding 81 events yielded similar estimates may indicate that spurious events were mainly observed at the periphery of the interaction network and hardly affected the behavior in the network's core. More generally, these results may assuage concerns over sensor data reliability (see \citealp{Elmer2019}).

\section{Discussion}
\label{sec:dis}

In summary, this paper extends the relational event framework to handle spurious events. In doing so, it offers applied researchers analyzing instantaneous interaction data a useful tool to explicitly account for measurement errors induced by spurious events or to investigate the robustness of their results against this type of error. Our proposed method controls for one explicit measurement error, namely that induced by spurious events. The simulation study showed that our approach can detect such false events and even yield correct results if they are not present. Still, we want to accentuate that numerous other types of measurement error may be present when one analyses relational events, which we disregard in this article. For instance, true events may be missing. These false negatives, e.g., unreported conflict events between different belligerents,  are difficult to tackle because of a lack of information.
 
We explicitly recommend the use of the REMSE as a method for checking robustness. When substantive knowledge suggests the presence of spurious events, the REMSE can be used to assess whether REM results hold when accounting for them. Spurious events may be common in datasets which come from sensors or are coded from journalistic sources, as discussed above, and more generally seem credibly present in data that are based on secondary sources instead of direct observation. Spurious events also occur, and may possibly be more influential, in data where relations are directed, the model we introduce accordingly also generalizes to directed event data. Especially for politically contentious data, where some events may be openly claimed to be false, the REMSE offers a possibility to adjudicate whether overall findings depend on such contested observations. But also where the data content is non-political, it is recommendable to check how common and influential false observations are. We provide replication code implementing the REMSE for this purpose.

When specifying the REMSE, two aspects require caution so that identifiability is ensured. First, given we know which events are spurious, our model simplifies to a competing risk model; thus, the identifiability issues discussed in \citet{Heckman1989} or \citet{TsiatisA1975} apply. For this reason, we presented our model under the assumption of independence between the \textsl{true-event} and \textsl{spurious-event} intensities. Second, the particular specification of the covariates might also affect the identifiability of the model. This may occur when one assumes complex dependencies of spurious events and exogenous covariates are unavailable, or the prior information about the coefficients is too weak. For the model specification employed in this article, this is not an issue due to the simple form of the \textsl{spurious-event} intensity as long as at least one exogenous or endogenous term has a nonzero effect on the \textsl{true-event} intensity. For more complex models, one may use multiple starting values of the data augmentation algorithm or formulate more informative priors for $\theta_1$ and possibly $\theta_0$.   

Our latent variable methodology can also be extended beyond the approach presented here. 
A straightforward refinement along the lines of \citet{Stadtfeld2017b} would be to include \textsl{windowed} effects, i.e., endogenous statistics that are only using history ranging into the past for a specific duration, or exogenous covariates calculated from additional networks to the one modeled. The first modification could also be extended to separable models as proposed in \citet{Fritz2020a}. A relatively simplistic version of the latter type of covariate was incorporated in Section \ref{sec:app_housing} to account for common friendships but more complex covariates are possible. This might be helpful, for instance, when we observe proximity and e-mail events between the same group of actors. 
Moreover, with minor adaptions, the proposed estimation methodology could handle some of the exogenous or endogenous covariates having nonlinear effects on the intensities. 

Finally, the framing of the simultaneous counting processes may be modified and their number extended. To better understand the opportunities our model framework entails, it is instructive to perceive the proposed model as an extension to the latent competing risk model of \citet{Alan2000} with two competing risks. 
For time-to-event data, one could thus employ an egocentric version\footnote{See \citealp{vu2011_2} for further information on egocentric models.} of our model for model-based clustering of general duration times, which could prove to be a valuable tool for medical applications. Or our proposed methodology could be conceived as a general tool to correct for additive measurement errors in count data and extend it to spatial data analysis to be used in settings described in \citep{Raleigh_Linke_Hegre_Karlsen_2010}. 

\appendix

\section{Definition of undirected network statistics}
\label{sec:annex_stat}

As REMs for undirected events are so far sparse in the literature, there are no standard statistics that are commonly used (one exception being \citealp{Bauer2019}). Thus we define all statistics based on prior substantive research \citep{rivera2010,Wasserman1994} and undirected statistics used for modeling static networks \citep{robins2007}. Generally, nondirected statistics have to be invariant to swapping the positions of actor $a$ and $b$. For the following mathematical definitions, we denote the set of all actors by $\mathcal{A}$. 

For degree-related statistics, we include the absolute difference of the degrees of actors $a$ and $b$: 
\begin{align*}
      s_{ab, Degree~Abs.}(\mathcal{H}(t)) &= |\sum_{h \in \mathcal{A}} (\mathbb{I}(N_{ah}(t^-)>0) + \mathbb{I}(N_{ha}>0)(t^-)) \\
    & \hspace{0.7cm}-  \sum_{h  \in \mathcal{A}} (\mathbb{I}(N_{bh}(t^-)>0) + \mathbb{I}(N_{hb}(t^-)>0))|,
\end{align*}
where $t^-$ is the point-in-time just before $t$. Alternatively, one might also employ other bivariate functions of the degrees as long as they are invariant to swapping $a$ and $b$, such as the sum of degrees. When simultaneously using different forms of degree-related statistics, collinearities between the respective covariates might severely impede the interpretation.     

To capture past dyadic behavior, one can include $N_{ah}(t^-)$ directly as a covariate. Since the first event often constitutes a more meaningful action than any further observed events between the actors $a$ and $b$, we additionally include a binary covariate to indicate whether the respective actors ever interacted before, leading to the following endogenous statistics: 
\begin{align*}
    s_{ab, Repition~Count}(\mathcal{H}(t)) &= N_{ah}(t^-) \\
    s_{ab, First~Repition}(\mathcal{H}(t)) &= \mathbb{I}(N_{ab}(t^-)>0).
\end{align*}

Hyperdyadic statistics in the undirected regime are defined as any type of triadic closure, where actor $a$ is connected to an entity that is also connected to actor $b$:
\begin{align*}
     s_{ab,Triangle} &=\sum_{h \in \mathcal{A}} \mathbb{I}(N_{ah}(t^-)>0) \mathbb{I}(N_{bh}(t^-)>0) + \\
      & \hspace{1.1cm}\mathbb{I}(N_{ha}(t^-)>0) \mathbb{I}(N_{bh}(t^-)>0) + \\
        & \hspace{1.1cm}\mathbb{I}(N_{ah}(t^-)>0) \mathbb{I}(N_{hb}(t^-)>0) + \\
          & \hspace{1.1cm}\mathbb{I}(N_{ha}(t^-)>0) \mathbb{I}(N_{hb}(t^-)>0)  \\
\end{align*}

Finally, actor-specific exogenous statistics can also be used to model the intensities introduced in this article. We denote arbitrary continuous covariates by $x_{a,cont} ~ \forall~ a \in \mathcal{A}$. On the one hand, we may include a measure for the similarity or dissimilarity for the covariate through: \begin{align*}
     s_{ab,Sim. ~ cont} &= |x_{a,cont}-x_{b,cont}| \\
     s_{ab,Dissim. ~ cont } &= \frac{1}{|x_{a,cont}-x_{b,cont}|}.
\end{align*}
For multivariate covariates, such as location, we only need to substitute the absolute value for any given metric, e.g., euclidean distance. In other cases, it might be expected that high levels of a continuous covariable result in higher or lower intensities of an event: 
\begin{align*}
     s_{ab,Sum~ cont} &= x_{a,cont} + x_{b,cont} \\.
\end{align*}
Which type of statistic should be used depends on the application case and the hypotheses to be tested. Categorical covariates, that we denote by $x_{a,cat} ~ \forall~ a \in \mathcal{A}$, can also be used to parametrize the intensity by checking for equivalence of two actor-specific observations of the variable: 
\begin{align*}
     s_{ab,Match~ cat} &= \mathbb{I}(x_{a,cat} = x_{b,cat}) \\.
\end{align*}
Besides actor-specific covariates also exogenous networks or matrices, such as $x_{Network} \in \mathbb{R}^{|\mathcal{A}|\times|\mathcal{A}|}$, can also be incorporated as dyadic covariates in our framework: 
\begin{align*}
     s_{ab,Dyad.~ Network} &= \frac{x_{Network, ab} + x_{Network, ba}}{2}, 
\end{align*}
where $x_{Network, ab}$ is the entry of the $a$th row and $b$th column of the matrix $x_{Network}$. Extensions to time-varying networks are straightforward when perceiving changes to them as exogenous to the modeled events \citep{Stadtfeld2017b}. 

\section{Mathematical derivation of \eqref{eq:bin}}
\label{sec:appenix_da}

For $m\in \{1, ..., M\}$, let $Y_{a_{m}b_{m},1}(t_{m}) =  N_{a_{m}b_{m},1} (t_{m}) - N_{a_{m}b_{m},1}(t_{m-1})$ be the increments of the latent counting process of true events between the time points $t_{m}$ and $t_{m-1}$, where we additionally define $t_0 = 0$ without the loss of generality. We observe $\mathcal{E}$, hence we can reconstruct the respective increment $Y_{a_{m}b_{m}}(t_{m}) = N_{a_{m}b_{m}}(t_{m}) - N_{a_{m}b_{m}}(t_{m-1}) = Y_{a_{m}b_{m},0}(t_{m}) + Y_{a_{m}b_{m},1}(t_{m})$, where $ Y_{a_{m}b_{m},0}(t_{m})$ is the increment of the spurious-event counting process. The second equality holds since by design the sum of increments of the processes counting the true and false events is the increment of the observed counting process, i.e. $N_{ab}(t) = N_{ab,0}(t)+N_{ab,1}(t)$. To sample from $Z_{m} | z_1, ..., z_{m-1},\mathcal{E}$, note that $Z_{m} = Y_{a_{m}b_{m},1}(t_{m}) | Y_{a_{m}b_{m}}(t_{m})$ holds. Heuristically, this means that if we know that one of the two \textsl{thinned} counting processes jumps at time $t_{m}$, the probability of the jump being attributed to $N_{a_{m}b_{m},1} (t)$ is the probability that the $m$th event is a true event. For the increments of the involved counting processes, we can then use the properties of the Poisson processes and the fact that the intensities are piecewise constant between event times to derive the following distributional assumptions $\forall~ m = 1, ..., M$: 
\begin{align}
   Y_{a_{m}b_{m},0}(t_{m}) | z_1, ..., z_{m-1},\mathcal{E}, \vartheta &\sim \text{Pois}\left(\delta_m\lambda_{a_{m}b_{m},0}(t_{m}| \mathcal{H}_0(t_{m}), \vartheta_0)\right) \label{eq:pro1} \\
   Y_{a_{m}b_{m},1}(t_{m}) | z_1, ..., z_{m-1},\mathcal{E}, \vartheta &\sim \text{Pois}\left(\delta_m\lambda_{a_{m}b_{m},1}(t_{m}, \mathcal{H}_1(t_{m}), \vartheta_1)\right) \label{eq:pro2} \\
    Y_{a_{m}b_{m}}(t_{m}) | z_1, ..., z_{m-1},\mathcal{E}, \vartheta &\sim \text{Pois}\bigg(\delta_m\big(\lambda_{a_{m}b_{m},0}(t_{m}|\mathcal{H}_0(t_{m}), \vartheta_0)   \label{eq:pro3}\\&\hspace{1.5cm}+ \lambda_{a_{m}b_{m},1}(t_{m}|\mathcal{H}_1(t_{m}), \vartheta_1)\big)\bigg),\nonumber
\end{align}
where we set $\delta_m = t_{m}- t_{m-1}$. We can now directly compute the probability of $Z_{m} = 1| z_1, ..., z_{m-1},\mathcal{E}, \vartheta$: 
\begin{align*}
    p(z_{m} = 1| z_1, ..., z_{m-1},\mathcal{E}, \vartheta) &= p(Y_{a_{m}b_{m},1}(t_{m})= 1 | Y_{a_{m}b_{m}}(t_{m}) = 1, z_1, ..., z_{m-1},\mathcal{E}, \vartheta) \\
    &= \frac{p(Y_{a_{m}b_{m},1}(t_{m})= 1, Y_{a_{m}b_{m}}(t_{m})= 1| z_1, ..., z_{m-1},\mathcal{E}, \vartheta)}{p(Y_{a_{m}b_{m}}(t_{m})= 1 | z_1, ..., z_{m-1},\mathcal{E}, \vartheta)} \\
   &= \frac{p(Y_{a_{m}b_{m},1}(t_{m})= 1, Y_{a_{m}b_{m},2}(t_{m})= 0| z_1, ..., z_{m-1},\mathcal{E}, \vartheta)}{p(Y_{a_{m}b_{m}}(t_{m})= 1 | z_1, ..., z_{m-1},\mathcal{E}, \vartheta)} \\
   &= \frac{p(Y_{a_{m}b_{m},1}(t_{m})= 1| z_1, ..., z_{m-1},\mathcal{E}, \vartheta)}{p(Y_{a_{m}b_{m}}(t_{m})= 1 | z_1, ..., z_{m-1},\mathcal{E}, \vartheta)} \\
   & \hspace{0.5cm}\times \frac{p(Y_{a_{m}b_{m},2}(t_{m})=0| z_1, ..., z_{m-1},\mathcal{E}, \vartheta)}{p(Y_{a_{m}b_{m}}(t_{m})= 1 | z_1, ..., z_{m-1},\mathcal{E}, \vartheta)} \\
   &= \frac{\lambda_{a_{m}b_{m},1}(t_{m}|\mathcal{H}_1(t_{m}, \vartheta_1))}{\lambda_{a_{m}b_{m},0}(t_{m}| \mathcal{H}_0(t_{m}), \vartheta_0)+ \lambda_{a_{m}b_{m},1}(t_{m}|\mathcal{H}_1(t_{m}), \vartheta_1 )}
\end{align*}
In the last row we plug in \eqref{eq:pro1} and \eqref{eq:pro2} for the probabilities in the numerators and  \eqref{eq:pro3} in the denominator to prove claim \eqref{eq:bin}. The calculation for $p(z_{m} = 0| z_1, ..., z_{m-1},\mathcal{E}, \vartheta)$ is almost identical to the one shown here.

\newpage
\bibliographystyle{chicago}
\bibliography{references}
    
\end{document}